# Electric field switching in a resonant tunneling diode electroabsorption modulator

José M. Longras Figueiredo, Charles N. Ironside and Colin R. Stanley

*Abstract*— The basic mechanism underlying electric field switching produced by a resonant tunneling diode (RTD) is analysed and the theory compared with experimental results; agreement to within 12% is achieved. The electroabsorption modulator (EAM) device potential of this effect is explored in an optical waveguide configuration. It is shown that a RTD-EAM can provide significant absorption coefficient change, via the Franz-Keldysh effect, at appropriate optical communication wavelengths around 1550 nm and can achieve up to 28 dB optical modulation in a 200 $\mu$m active length device. The advantage of the RTD-EAM, over the conventional reversed biased pn junction EAM, is that the RTD-EAM has in essence an integrated electronic amplifier and therefore requires considerably less switching power.

## I. Introduction

WITH the steady improvement of high precision growth techniques for semiconductor layers, in particular Molecular Beam Epitaxy (MBE), there has been a renaissance in tunneling devices for electronic applications. Compared to previous attempts to produce tunneling devices, high precision growth now gives much more control over device characteristics which are crucially dependent on layer thickness and tunneling devices are now being considered as memory devices [1] and a new logic family has been proposed [2]. Furthermore, it has been demonstrated that III-V semiconductor tunneling devices can be integrated with silicon CMOS technolgy and that tunneling devices can be driven by CMOS logic levels [3]. The physics and progress in electronic applications of Resonant Tunneling Diodes (RTDs) have recently been reviewed in [4].

Optoelectronics has provided perhaps the most impressive example of high precision growth for a tunneling device with the invention of the quantum cascade laser [5] in which population inversion between subbands in quantum wells is produced by carefully engineered tunneling. Simpler optoelectronic device structures essentially based on double barrier resonant tunneling diodes (DBRTD) have also been used in various applications; these include photodetectors at optical communication wavelengths [6], mid-infrared wavelengths [7] and, closely related to the work presented here, optical modulators [8][9][10].

In this paper we report on the application of a double barrier resonant tunneling diode (RTD) to electroabsorption modulator (EAM) devices and in particular the electric field switching which is the basis of the operation of the device. We have been investigating this as an alternative to conventional EAM devices which are currently employed in optical communication systems and where the electric field is applied and switched by employing a reversed biased pin diode. The key advantage of the RTD-EAM over the conventional pin-EAM (for a recent review of conventional pin-EAMs see [11]) is that the RTD-EAM can provide electrical gain over a wide bandwidth, and thereby achieve a low-drive voltage and high speed operation.

## II. Principle of operation of the RTD-EAM

Essentially, the RTD-EAM is a unipolar device which consists of a DBRTD embedded in an optical waveguide. The presence of the DBRTD within the waveguide core introduces high non-linearities in the current-voltage (I-V) characteristic of the unipolar waveguide. A typical I-V characteristic of a RTD-EAM is shown in Fig. 1 (the physics which gives rise to this type of I-V has been previously explained [4]).

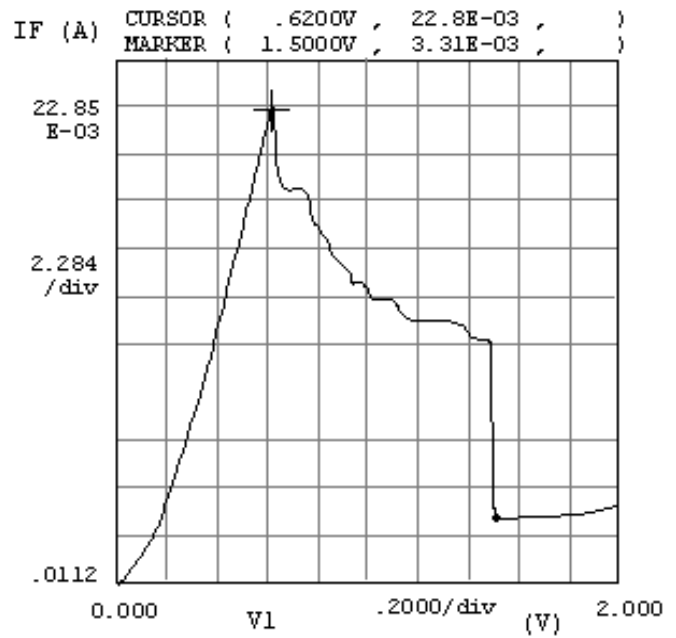

Fig. 1. I-V characteristic of 2 $\mu$m × 100 $\mu$m active area RTD-EAM.

The operation of the RTD-EAM is based on a non-uniform electric field distribution across the waveguide induced by the DBRTD which becomes strongly dependent on the bias voltage. When the current decreases from the

J. M. Longras Figueiredo is with the Department of Physics, Faculdade de Ciências e Tecnologia, Universidade do Algarve, 8000-117 Faro, Portugal. E-mail: jlongras@ualg.pt.
C. N. Ironside and C. R. Stanley are with the Department of Electronics and Electrical Engineering, University of Glasgow, Glasgow G12 8LT, United Kingdom.

peak to the valley there is an increase of the electric field across the waveguide core. The electric field enhancement in the depleted spacer layer causes the Franz-Keldysh absorption band-edge shift to lower energy which is responsible for the electroabsorption effect.

In a conventional EAM the electric field is applied by reverse biasing a pn diode that shifts the absorption band-edge of the depleted region to lower energy. The key difference with the RTD-EAM is that the tunneling characteristics of the double barrier RTD are employed to switch the electric field across the waveguide collector depleted region. Therefore, a small high frequency ac signal (<1 V) can induce high speed switching producing substantial modulation of light at photon energy slightly lower than the waveguide band-gap energy.

The RTD-EAM is implemented in a ridged channel unipolar waveguide configuration lying on top of the substrate, Fig. 2(a), and its structure resembles the quantum well injection transit time (QWITT) diode proposed by Kesan et al [12] as illustrated in Fig. 2(b). In fact, the RTD-EAM has many similarities to transit-time devices as the tunneling current acts as an injection source to the collector depleted region [13].

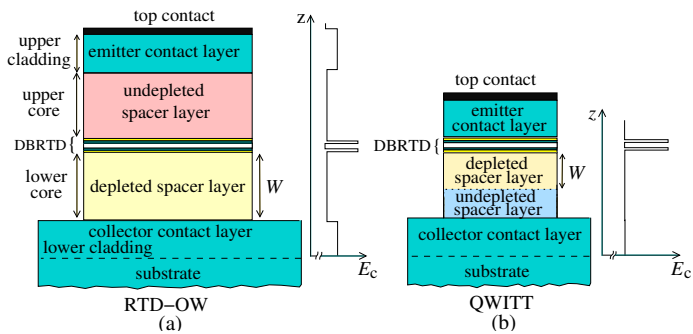

Fig. 2. RTD-EAM and QWITT schematic structures.

In essence, the RTD-EAM is a DBRTD current switch in series with a resistor, except that the speed of response is limited by the electron transit time across the collector depleted spacer layer; if one assumes an electron saturation velocity, $v_{sat}$, of $10^7$ cm/s and a depletion region width $W \sim 500$ nm (see Fig. 4), the electron transit time is 5 ps.

The physical mechanism by which the current drop is converted into an electric field enhancement is as follows. When the bound state of the DBRTD quantum well is above or aligned with the emitter conduction band energy minimum (see diagram of Fig. 3) the electron transmission is high and the carriers can easily tunnel through the bound state with little free carrier depletion in the collector region. The applied voltage is dropped mainly across the DBRTD and the electric field gradient in the collector spacer layer is small because the spacer layer is not strongly depleted. This corresponds to the transmissive state (on-state) of the modulator (during operation the RTD-EAM is dc biased slightly below the peak voltage). Once the applied voltage is increased from the peak to the valley, the DBRTD bound state is pulled below the emitter conduction band energy minimum, as depicted in Fig. 3, and the electrons can no longer tunnel through using the bound state. The current through the device drops giving rise to an increase of positive space charge in the collector region; a substantial part of the terminal voltage is now dropped across the collector spacer layer. As a consequence, the magnitude of the electric field in the collector spacer layer increases: this is the non-transmissive state of the modulator. To summarise, the peak-to-valley current drop produces an increase in the magnitude of the electric field across the waveguide core collector region. This causes the broadening of the waveguide absorption band-edge through the Franz-Keldysh effect to longer wavelengths, which in turn leads to an increase of the optical absorption coefficient of photons possessing energy slightly lower the waveguide band-edge energy.

In Fig. 3 we show the energy band diagram in the RTD-EAM at the valley voltage where the applied voltage is dropped mainly across the depleted region of the waveguide core. To determine the magnitude of the electric field change in the depleted core induced by peak-to-valley switching, the injection region (the emitter and the DBRTD, Fig. 3) is decoupled from the depletion region since its characteristics should not depend strongly on the collector spacer layer [13]. Our analysis follows the QWITT model (see also [14]).

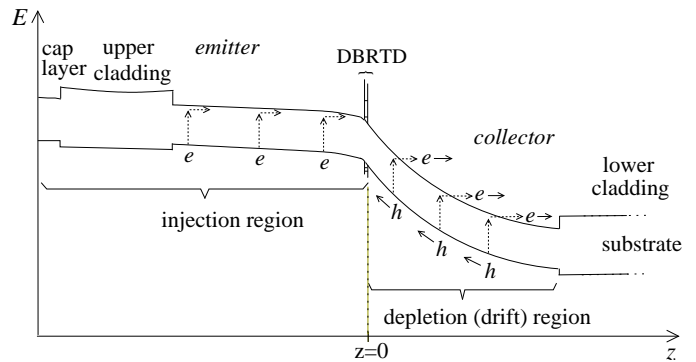

Fig. 3. Schematic diagram of the energy bands in a RTD-EAM at the valley voltage as a function of distance: upper curve is the lowest conduction band energy and the lower curve represents the highest valence band energy.

In the analysis, the electric field across the depleted spacer layer (or drift region) is assumed to be high enough to cause the injected electrons to traverse the drift region at a constant saturation velocity, $v_{sat}$. Quantitatively, the magnitude of the electric field in the drift region as function of position $\mathcal{E}(z)$ at a constant current density $J$ can be obtained from the one dimensional Poisson's equation as follows:

$$\frac{\partial \mathcal{E}}{\partial z} = \frac{e}{\epsilon}\left(N_d - \frac{J}{ev_{sat}}\right), \qquad (1)$$

where $z$ is the distance, $e$ is the electronic charge, $\epsilon$ is the core permittivity and $N_d$ is the background doping density in the depleted spacer layer. Integrating, the electric field across the depletion region becomes

$$\mathcal{E}(J;z) = \mathcal{E}_0(J) - \frac{e}{\epsilon}\left(N_d - \frac{J}{ev_{sat}}\right)\cdot z, \quad (2)$$

where $\mathcal{E}_0(J)$ represents the electric field at the injection plane $z = 0$, the boundary between the DBRTD and the depleted spacer layer (see Fig. 3). The change in voltage across the drift region due to peak-to-valley switching is given by:

$$\Delta V_d = \int_0^W [\mathcal{E}(J_v;z) - \mathcal{E}(J_p;z)]\,dz = \Delta\mathcal{E}_0 W + \frac{W^2}{2\epsilon v}(J_v - J_p), \quad (3)$$

where $\Delta\mathcal{E}_0 = \mathcal{E}_{0,J_v} - \mathcal{E}_{0,J_p}$ is the change in the electric field at $z = 0$, between the valley and peak points. The width of the depleted spacer layer, $W$, is assumed to remain constant before and after current switching, as is $N_d$. ($W$ is defined by the thickness of the low doped layer in the collector region of the RTD-EAM.) We can re-arrange the above equation to obtain the peak-to-valley electric field enhancement at the injection plane

$$\Delta\mathcal{E}_0 = \frac{\Delta V_d}{W} + \frac{W}{2\epsilon v}\Delta J_{p-v}. \quad (4)$$

The changes in voltage, $\Delta V_d$, and current, $\Delta J_{p-v} = J_p - J_v$, are found from the I-V characteristic of the RTD-EAM.

The analysis of Eq. 2 points out an important consideration limiting the electric field switching: the background doping of the depleted spacer layer (drift region), $N_d$, must be greater than $J_p/ev_{sat}$; if this conditions is not observed then the electric field increases with the distance and can exceed the breakdown value. ($J_p$ is essentially determined by the DBRTD structure.) Experimentally, we observed in the InP based RTD-EAM that $N_d = 2 \times 10^{16}$ cm$^{-3}$ gave unreliable devices subject to catastrophic breakdown whereas $N_d = 5 \times 10^{16}$ cm$^{-3}$ produced reliable devices. Also, doping across the structure, specially in the cladding layers, has to be kept as high as possible to minimise the device series resistance; however, to minimise optical loss due to free carrier losses it is desirable to keep $N_d$ as small as possible. A compromise between the required electrical and optical properties is needed.

III. WAFER DESIGN AND GROWTH, FABRICATION AND PACKAGING

The RTD-EAM is a unipolar optical waveguide containing a DBRTD; the DBRTD is employed to switch the electric field developed across the waveguide collector region as described above. The optical waveguide configuration ensures a larger interaction volume between the active region of the device (RTD depletion region) and the optical mode, thereby ensuring a larger modulation depth for a given applied field. The wavelength of operation is set by the band-gap of the material employed in the active region (waveguide core) of the device. Our initial devices used GaAs in the active region (see [8][9]) and operated at 900 nm, subsequently In$_{0.53}$Ga$_{0.42}$Al$_{0.05}$As was employed to shift the wavelength of operation to 1550 nm (InGaAlAs was used because it is a convenient semiconductor alloy for MBE growth.)

The layer design for the InGaAlAs RTD-EAM device is shown in Fig. 4. The RTD-EAM wafers were grown by Molecular Beam Epitaxy in a Varian Gen II system on a InP substrate. The waveguide core was formed by two moderately doped (Si: $5\times10^{16}$ cm$^{-3}$) In$_{0.53}$Ga$_{0.42}$Al$_{0.05}$As layers 500 nm thick (absorption band edge around 1520 nm and refractive index of 3.56) each side of the DBRTD (2 nm thick AlAs barriers and 6 nm thick In$_{0.53}$Ga$_{0.47}$As quantum well). The upper cladding layer of the optical waveguide consisted of a 300 nm In$_{0.52}$Al$_{0.48}$As layer heavily doped (Si: $2\times10^{18}$ cm$^{-3}$). The contact layer was a In$_{0.53}$Ga$_{0.47}$As layer $\delta$-doped for the formation of non-alloyed Au-Ge-Ni ohmic contacts.

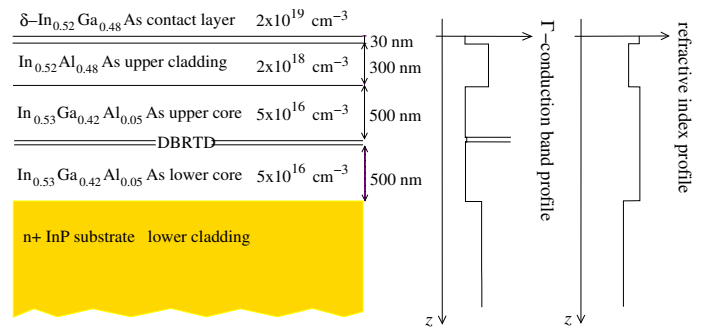

Fig. 4. InGaAlAs/InP wafer structure, Γ-valley and refractive index profiles.

Ridge waveguides (2 to 6 $\mu$m wide) and large-area mesas on each side of the ridges were fabricated by wet-etching. Ohmic contacts (100 to 400 $\mu$m long) were deposited on top of the ridges and mesas. The waveguide width and the ohmic contact length define the device active area. A SiO$_2$ layer was deposited, and access contact windows were etched on the silica over the ridge and the mesa electrodes, Fig. 5, allowing contact to be made through high frequency bonding pads (coplanar waveguide transmission line, CPW). Figure 5 shows the layout of the RTD-EAM chip and Fig. 6 is an annotated photograph showing the top view of a finished RTD-EAM die.

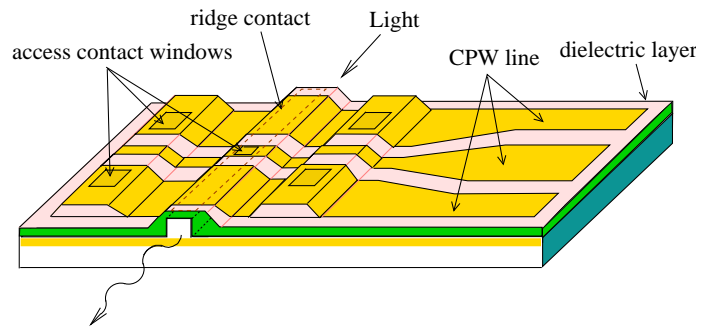

Fig. 5. Schematic of the RTD-EAM.

After cleaving, the devices were die bonded on packages allowing light to be coupled into the waveguide by a mi-croscope objective end-fire arrangement. The details of the fabrication procedure and device packaging can be found in [14].

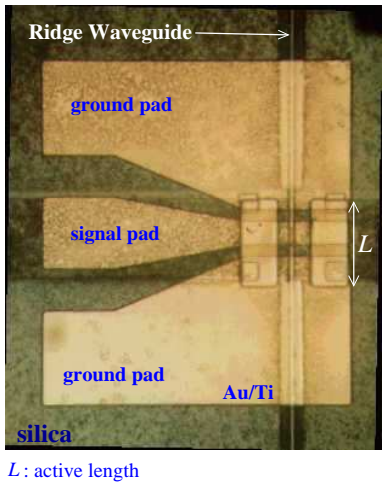

Fig. 6. Top view of a RTD-EAM die (430 $\mu$m $\times$ 500 $\mu$m with $L$=100 $\mu$m), showing the CPW contact-pad/transmission line.

## IV. Results

Here we report on the electrical and the optical characterisation of the InGaAlAs/InP RTD-EAM. The electrical characterisation was concerned with dc measurements of the current-voltage curve and the optical characterisation was concerned with low frequency modulation and background loss measurements.

The dc electrical characteristics of the packaged devices were measured using a HP 4145 parametric analyser. Figure 7 shows two typical I-V curves for devices of different active lengths and widths. The peak voltage rises as the active area increases. Typical $4\times 200$ $\mu$m$^2$ active area RTD-EAMs show peak current densities up to 18 kAcm$^{-2}$ and peak-to-valley current ratios (PVCRs) of around 4, with valley-to-peak voltage difference, $\Delta V_{v-p}$, of 0.8 V and peak to valley current density change up to $\Delta J_{p-v}$= 13 kAcm$^{-2}$. Some devices showed larger PVCR and up to 7 could be achieved (see Fig. 1).

An interesting feature of the I-V curves is that they are not completely anti-symmetric, the peak voltage in the reverse region is smaller in magnitude compared to the positive voltage side and indeed it was frequently the case that devices operated in the forward voltage showed catastrophic electrical breakdown. This behaviour may be related to the unsymmetric nature of the layer structure and the different electrical characteristics of the InAlAs alloy, forming the upper cladding, compared to the InP substrate which acts as the lower cladding. In the optical experiments described below devices were operated in the negative voltage side of the I-V curve, i.e., electrons flowing towards the substrate.

Optical characterisation of the InGaAlAs RTD-EAMs employed a diode laser, tuneable in the wavelength range 1480-1580 nm. End-fire and fibre coupling into the semi-conductor waveguide were employed. The waveguide was not single mode but it was possible to excite individual modes with a single mode fiber. By tuning the laser and measuring the throughput the waveguide transmission spectra was obtained; the Fabry-Perot etalon method was employed to calibrate the spectra by measuring the loss at a particular wavelength. For zero applied field, the absorption coefficient at 1565 nm was found to be 8.2 cm$^{-1}$. The change in the absorption spectrum was measured at various points in the I-V curve of the RTD-EAM, as shown in Fig. 8(a). The absorption change induced by peak-to-valley switching, $\Delta\alpha = \alpha_v - \alpha_p$, is presented in Fig. 8(b). Figure 8(c) summarises the results of these experiments, showing the degree of electroabsorption characterised by the change in optical absorption coefficient induced by the peak-to-valley transition $\Delta\alpha$, over the absorption coefficient at the peak bias $\alpha_p$, $\Delta\alpha/\alpha_p$.

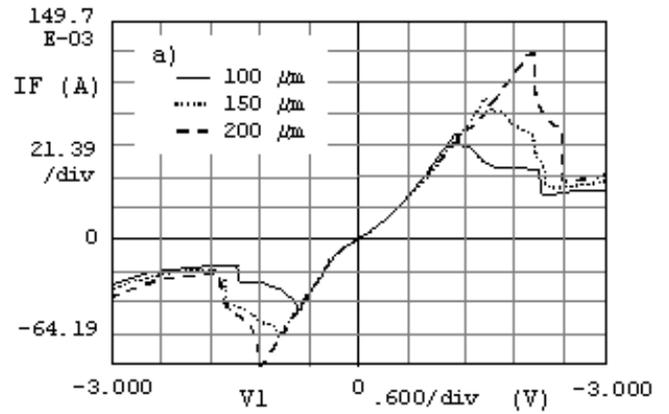

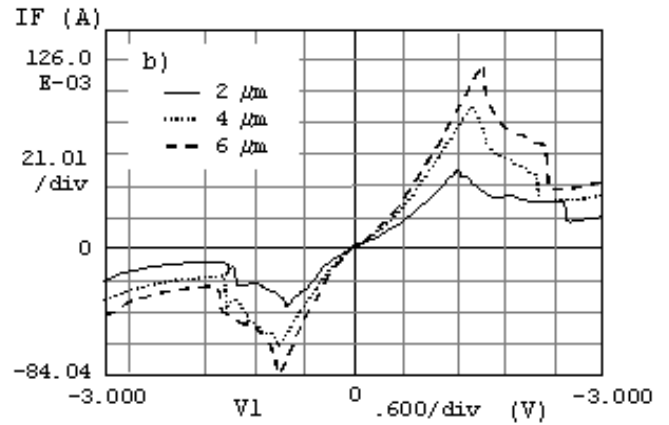

Fig. 7. RTD-EAM I-V characteristics with active length [100, 150 and 200 $\mu$m; 4 $\mu$m wide, a)] and width [2, 4 and 6 $\mu$m; 150 $\mu$m long, b)], respectively, as parameter.

Preliminar electroabsorption modulation results have been previously reported [10]; in summary, devices with $4 \times 200$ $\mu$m$^2$ active area showing the highest PVCR and largest $\Delta V_{v-p}$, when dc biased to the optimum operating point, had a maximum modulation depth of 28 dB at around 1565 nm [10]. Typical $4 \times 200$ $\mu$m$^2$ active area de-vices showed a modulation depth of around 20 dB in the wavelength range 1560-1567 nm, with propagation loss in the transmissive state estimated to be $\sim$5 dB [14]. The quiescent power required to bias the device in the on-state (near the switching point, $V_p$), is around 30 mW. Furthermore, in collaboration with other groups [15] we have shown that with these devices it is possible to achieve a modulation of 5 dB for a voltage change of 1 mV. We estimate that a 10 dB of modulation can be achieved for a power of a few milliwatt.

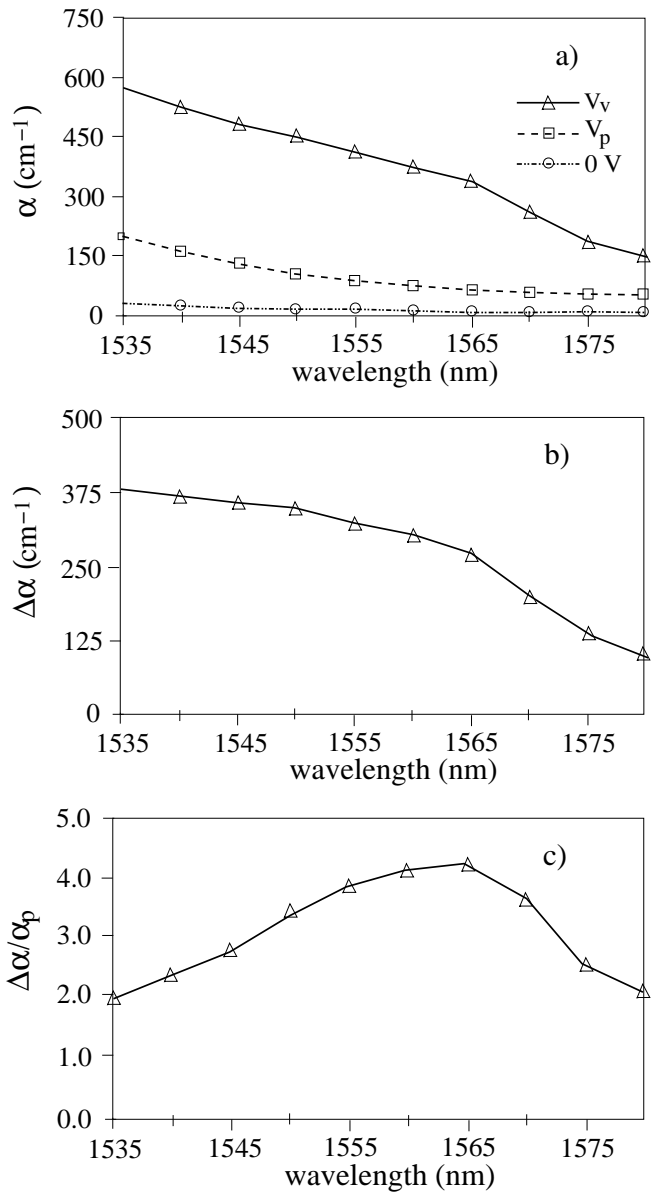

Fig. 8. Typical RTD-EAM absorption spectra characterisation: a) absorption, $\alpha$ at zero, at the peak and at the valley voltages; b) absorption change, $\Delta\alpha = \alpha_v - \alpha_p$, where $\alpha_v$ and $\alpha_p$ represent the absorption at the valley and peak points, respectively; and c) change in the absorption coefficient over the absorption coefficient at peak voltage, $\Delta\alpha/\alpha_p$.

## V. Comparison with theory

Considering the $4 \times 200$ $\mu$m$^2$ active area InGaAlAs RTD-EAM devices with measured values of $\Delta V_{v-p}$ around 0.8 V and $\Delta J_{p-v}$ =13 kAcm$^{-2}$, taking $\epsilon$=14$\epsilon_0$, $v_{sat}$=1 $\times$ 10$^7$ cm/s [16], we find an electric field enhancement $\Delta\mathcal{E}_0$=43 kV/cm, Eq. 4. (In Eq. 4, $W$ corresponds to the width of low doped layer on the collector side of the RTD; in this present structure is 500 nm.)

In the calculation of the absorption band-edge shift a uniform electric field across the depleted region is assumed, and any shift due to thermal effects as a consequence of the current flow and the peak voltage is neglected. Assuming $\mathcal{E}_0(J_v) \gg \mathcal{E}_0(J_p)$, $\Delta\mathcal{E}_0 \cong \mathcal{E}_0(J_v)$, the shift in the InGaAlAs waveguide transmission spectrum due to electric field enhancement as a result of the Franz-Keldysh effect, is given approximately by [8]

$$\Delta\lambda_g \cong \frac{\lambda_g^2}{hc}\left(\frac{e^2h^2}{8\pi^2 m_r}\right)^{\frac{1}{3}} \Delta\mathcal{E}^{\frac{2}{3}}. \quad (5)$$

where $m_r$ is the electron-hole system reduced effective mass, $h$ is the Planck's constant, $c$ is the light velocity, $e$ is the electron charge, and $\lambda_g$ is the wavelength corresponding to the waveguide transmission edge at zero bias.

To calculate the band-edge shift we use the bandgap wavelength of In$_{0.53}$Ga$_{0.42}$Al$_{0.05}$As, $\lambda_g$=1520 nm, and the following effective masses values (appropriate for the In$_{0.53}$Ga$_{0.42}$Al$_{0.05}$As alloy), $m_e = 0.046 m_0$ and $m_{hh} = 0.5 m_0$ (only the lowest electron-to-heavy-hole transitions are considered). From the above equation, we obtain $\Delta\lambda_g = 46$ nm compared to the observed value $\Delta\lambda_g = 43$ nm, indicating an actual electric field of 38 kV/cm. This provides evidence that Eq. 4 predicts, as a first approximation, the magnitude of the electric field enhancement to 12% accuracy. We therefore conclude that our model gives a good estimate of the switched electric field $\Delta\mathcal{E}_0$ and thereby the band-edge shift can be approximately calculated.

The RTD has been described as the fastest purely electronic device and oscillation at up to 712 GHz has been reported from an RTD device [17]. Streak camera studies of our GaAs RTD-EAM demonstrated 30 ps pulses of light and self-oscillation at up to 16 GHz which analysis of the design suggested it was limited by the electronic packaging [9][14]. A similar streak camera investigation of the InAlGaAs RTD-EAM high-speed characteristics could not be undertaken because of the low radiant sensivity of the streak tube at 1550 nm. The next stage in the development of the InAlGaAs RTD-EAM will be to employ other types of detector for an investigation of its high speed operation. These studies will be undertaken after further optimisation of the design of the device and the package.

## VI. Conclusions

We have presented a study of the basic mechanism of electric field switching with a RTD and we have demonstrated how this electric field switching can be employed in an electroabsorption modulator device configuration to obtain optical modulation at 1550 nm. The analysis of the electric field switching is based on the QWITT model [12][13] which explicitly takes account of the electric field across the depleted spacer layer. In the RTD-EAM it is the band-edge shift in the depleted spacer layer, via the Franz-Keldysh effect, which gives rise to the observed optical modulation; the optical waveguide configuration is employed to confine light in the depleted spacer layer. The value of the switched electric field was deduced from the absorption band-edge shift and the model was shown to give an estimate to within 12% of this switched electric field. The model assumes a uniform electric field across the depleted region but it could be further refined by taking account of the graded electric field across the depleted spacer layer.

The change in the optical absorption coefficient induced by the peak-to-valley transition over the absorption coefficient at the peak current, $\Delta\alpha/\alpha_p$, associated with the switch in electric field shows a maximum value of 4 at a wavelength of 1565 nm. This effect was employed (combined with an applied voltage swing slightly larger than $\Delta V_{v-p}$) to obtain a maximum modulation of 28 dB at 1565 nm in a device with an active region length of 200 $\mu$m. The magnitude of the switched electric field in the RTD-EAM was found to be around $\Delta\mathcal{E}_0$=43 kV/cm which compared with >100 kV/cm obtainable with a pin-EAM [11].

The insertion loss due to the absorption coefficient in the transmissive state, $\alpha_p$, is a material property which depends on the electric field value at the on-state operating voltage [11]. The insertion loss caused by the waveguide-fiber mode mismatch and the reflections at the facets have the same magnitude as conventional EAM's. The reduction in the on-state voltage and the improvement of the guiding characteristics of the waveguide, which we believe can be achieved in an optimised device, will lead to a smaller $\alpha_p$, decreasing the insertion loss to similar levels of conventional EAM's.

The expected quiescent power dissipation of 30 mW in the on-state in typical applications can be regarded as dc power supply. The conventional EAM chip requires less quiescent drive power (it depends on the transmissive state bias electric field). However, it needs a rf amplifier to drive it. The RTD-EAM provides on-chip electrical amplification which can be employed to substantially reduce the power required from the high frequency (rf) applied data signal and thereby remove the requirement of an external rf amplifier. In addition, the electrical characteristics of the RTD are well suited for digital modulation and can be combined with recent developments in RTDs for purely electronic applications to develop a new high-speed digital technology. Driving many of these developments is the intrinsic high speed of the tunneling process which has already been demonstrated to operated at over 700 GHz and could be harnessed in a combination optoelectronic and electronic devices to provide the high speed communication, memory and processing required by the next generation of information technology. However, it should also be noted that the operating parameters of tunneling devices are sensitive to device dimension on a nanometer scale and even with the steady improvement in epitaxial growth techniques there are still significant challenges ahead in making this a manufacturable technology [18].

## References

[1] J. P. A. Van Der Wagt, "Tunneling based SRAM," *Proc. IEEE*, vol. 87, pp. 571-595, 1999.

[2] R. H. Mathews, J. P. Sage, T.C.L. G. Sollner, S. D. Calawa, C. Chen, L. J. Mahoney, P. A. Maki, and Karen M. Molvar, "A new RTD-FET Logic Family," *Proc. IEEE*, vol. 87, pp. 596-605, 1999.

[3] J. I. Bergman, J. Chang, Y. Joo, B. Matinpour, J. Lasker, N. M. Joerst, M.A. Brooke, B. Brar, and E. BEam, III, "RTD/CMOS nanoelectronic circuits: thin-film InP-based resonant tunneling diodes integrated with CMOS circuits," *IEEE Elect. Device Lett.*, vol. 20, pp. 119-122, 1999.

[4] H. Mizuta, and T. Tanoue, *Physics and Applications of Resonant Tunneling Diodes*, Cambridge University Press, 1995.

[5] J. Faist, F. Capasso, D. L. Sivco, C. Sitori, A. L. Hutchinson, and A. Y. Cho, "Quantum Cascade Laser," *Science*, vol. 264, pp. 533-556, 1994.

[6] T. S. Moise, Y.-C. Kao, C. L. Goldsmith, C. L. Schow, and J. C. Campbell, "High-speed resonant tunneling diodes photodiodes with low switching energy," *IEEE Photon. Technol. Lett.*, vol. 9, pp. 803-805, 1997.

[7] C. Mermelstein, and A. Sa'ar, "Intersubband photocurrent from double barrier resonant tunneling structures," *Supperlattices and Microstructures*, vol. 19, pp. 375-382, 1996.

[8] S. G. McMeekin, M. R. S. Taylor, B. Vögele, C. R. Stanley, and C. N. Ironside, "Franz-Keldysh effect in optical waveguide containing a resonant tunneling diode," *Appl. Phys. Lett.*, 65, pp. 1076-1078, 1994.

[9] J. M. L. Figueiredo, C. R. Stanley, A. R. Boyd, C. N. Ironside, S. G. McMeekin, and A. M. P. Leite, "Optical Modulation in a resonant tunneling relaxation oscillator," *Appl. Phys. Lett.*, vol. 74, pp. 1197-1199, 1999.

[10] J. M. L. Figueiredo, C. R. Stanley, A. R. Boyd, C. N. Ironside, S. G. McMeekin, and A. M. P. Leite, "Optical Modulation at around 1550 nm in InGaAlAs optical waveguide containing a InGaAs/AlAs resonant tunneling diode," *Appl. Phys. Lett.*, vol. 75, pp. 3443-3445, 1999.

[11] K. Wakita, *Semiconductor Optical Modulators*, Kluwer Academic, 1998.

[12] V. P. Kesan, D. P. Neikirk, B. G. Streetman, and P. A. Blakey, "The Influence of Transit-Time Effects on the Optimum Design and Maximum Oscillation Frequency of Quantum Well Oscillators," *IEEE Trans. Electron Devices*, vol. 35, pp. 405-413, 1988.

[13] S. Yngvesson, *Microwave Semiconductor Devices*, Kluwer Academic Publishers, 1991.

[14] J. M. L. Figueiredo, *Optoelectronic Properties of Resonant Tunnelling Diodes*, University of Porto, Portugal, Ph.D. thesis, 2000.

[15] J. M. L. Figueiredo, A. M. P. Leite, C. N. Ironside, C. R. Stanley, S. G. McMeekin, K. Bouris, and D. G. Moodie, "Resonant tunneling diode electroabsorption waveguide modulator operating at around 1550 nm," *Tech. Dig. Conf. Laser Electro-Optics (CLEO 2000)*, pp. 596-597, 2000.

[16] A. F. J. Levi, "Nonequilibrium electron transport in heterojunction bipolar transistors," Chapter 4 in *Growth, Processing and Applications InP HBTs*, Editors B. Jalali and S. J. Pearton, Artech House, 1995.

[17] E. R. Brown, J. R. Soderstrom, C. D. Parker, L. J. Mahoney, K.M. Molvar, and T. C. McGill, "Oscillations up to 712 GHz in InAs/AlSb resonant-tunneling diodes," *Appl. Phys. Lett.*, vol. 85, pp. 2291-2293, 1991.

[18] V. A. Wilkinson, M. J. Kelly, and M. Carr, "Tunnel devices are not yet manufacturable," *Semiconductor Sci. Technol.*, vol. 12, pp. 91-99, 1997.

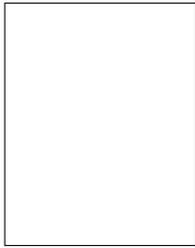
**José M. Longras Figueiredo** was born in Barcelos, Portugal. He received the B.Sc. in Physics (Optics and Electronics) in 1991 from the University of Porto. His final year project on holographic optical elements aberration correction with an intermediate computer generated hologram was completed under the ERAMUS Programme at the Vrije Universiteit Brussels. From 1992 to 1995 he was actively involved on the RACE II Project POPCORN (Polymeric Passive Components Research for Optical Networks). His M.Sc. in Optoelectronics and Lasers was completed in 1995 with a thesis on wavelength division multiplexing integrated optical devices. From 1995 to 1999 he was at the Department of Electronics and Electrical Engineering of the University of Glasgow, Scoltand, as a Ph.D. student working on optoelectronic properties of resonant tunneling diodes. His current research interests include the design and characterisation of optoelectronic devices.

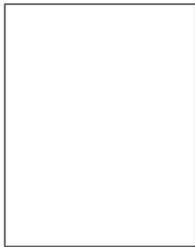
**Charles N. Ironside** was born in Aberdeen, Scotland. He completed his first degree in Physics at Heriot-Watt University, Edinburgh in 1974 and was award a first class honours BSc. His PhD was completed in 1978 and was on a type of tuneable semiconductor laser, the Spin-Flip Raman Laser. His PhD was supervised by Prof. S. D. Smith at the Physics Department of Heriot-Watt University. From 1978-1984 he worked as a post-doctoral research assistant at the University of Oxford. He first worked in the Inorganic Chemistry department on time resolved spectroscopy of solids and on energy transfer mechanisms between luminescent ions. He then moved to the Clarendon Laboratory to work on time-resolved spectroscopy of solids on a picosecond timescale. He worked on ultrafast effects in semiconductors. In 1984 he moved to the Department of Electronics and Electrical Engineering, University of Glasgow. He has published more than 150 research papers and presently his research interests include ultrafast all-optical switching in semiconductor waveguides, monolithic modelocking of semiconductor lasers and ultrafast optoelectronic modulators employing Resonant Tunneling Diodes and Quantum Cascade Lasers.

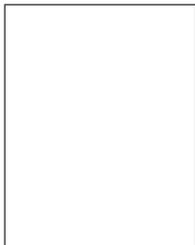
**Colin R. Stanley** was born in Tunbridge Wells, England in 1945. He graduated with a 1st class honours degree in Electronic Engineering from the University of Sheffield in 1966, and gained his Ph.D. from the University of Southampton in 1970 for research into the nonlinear optical properties of tellurium. From 1970-1972, he was I.C.I. Research Fellow at the University of Southampton, continuing work on optical parametric amplifiers and oscillators. He spent four months in 1971 at the Centre Nationale d'Etudes des Telecommunications at Bagneux, Paris as a Visiting Scientist. He joined the Department of Electronics and Electrical Engineering at the University of Glasgow in 1972 as a Research Fellow, and was subsequently appointed Lecturer (1974), Senior Lecturer (1982), Reader (1989) and Titular Professor (1992). Professor Stanley was a Visiting Scientist at Cornell University (USA) for eight months in 1982, and spent six months from October 1997 on industrial secondment with Motorola in East Kilbride, under a scheme financed by the Royal Academy of Engineering. He is a member of the EPSRC Functional Materials College, the Institution of Electrical Engineers and the Institute of Physics. He has been author or co-author of over 90 papers on MBE and related topics.